\documentclass[twocolumn,showpacs,superscriptaddress,preprintnumbers,amsmath,
amssymb,prl]{revtex4}

\usepackage{graphicx}
\usepackage{dcolumn}
\usepackage{bm}
\usepackage{hyperref}

\begin{document} 

\title{Homopolar bond formation in ZnV$_2$O$_4$ close to a metal-insulator transition}

\author{V. Pardo}
 \email{vpardo@usc.es}
\affiliation{
Departamento de F\'{\i}sica Aplicada, Universidad
de Santiago de Compostela, E-15782 Santiago de Compostela,
Spain
}
\affiliation{
Instituto de Investigaciones Tecnol\'ogicas, Universidad de Santiago de
Compostela, E-15782, Santiago de Compostela, Spain
}

\author{S. Blanco-Canosa}
\affiliation{
Departamento de Qu\'{i}mica-F\'{\i}sica, Universidad
de Santiago de Compostela, E-15782 Santiago de Compostela,
Spain
}

\author{F. Rivadulla}
\affiliation{
Departamento de Qu\'{i}mica-F\'{\i}sica, Universidad
de Santiago de Compostela, E-15782 Santiago de Compostela,
Spain
}

\author{D.I. Khomskii}
\affiliation{
II. Physikalisches Institut, Universit\"at zu K\"oln, Z\"ulpicher Str. 77, D-50937 K\"oln, Germany}

\author{D. Baldomir}
\affiliation{
Departamento de F\'{\i}sica Aplicada, Universidad
de Santiago de Compostela, E-15782 Santiago de Compostela,
Spain
}
\affiliation{
Instituto de Investigaciones Tecnol\'ogicas, Universidad de Santiago de
Compostela, E-15782, Santiago de Compostela, Spain
}

\author{Hua Wu}
\affiliation{
II. Physikalisches Institut, Universit\"at zu K\"oln, Z\"ulpicher Str. 77, D-50937 K\"oln, Germany}

\author{J. Rivas}
\affiliation{
Departamento de F\'{\i}sica Aplicada, Universidad
de Santiago de Compostela, E-15782 Santiago de Compostela,
Spain
}

\begin{abstract}

Electronic structure calculations for spinel vanadate ZnV$_2$O$_4$ show that partial electronic delocalization in this system leads to structural instabilities. These are a consequence of the proximity to the itinerant-electron boundary, not being related to orbital ordering. We discuss how this mechanism naturally couples charge and lattice degrees of freedom in magnetic insulators close to such a crossover. For the case of ZnV$_2$O$_4$, this leads to the formation of V-V dimers along the [011] and [101] directions that readily accounts for the intriguing magnetic structure of ZnV$_2$O$_4$.

\end{abstract}


\maketitle


The transition between an antiferromagnetic (AF) insulator and a paramagnetic metal is associated to some of the most intriguing experimental observations in solid state physics. Unconventional forms of superconductivity, electronic phase separation, and a variety of non-Fermi liquid behavior are well-known examples. Although not completely understood, it is however becoming clear that a strong coupling between charge, spin, orbital
 and lattice degrees of freedom is an essential ingredient to explain many of these phenomena. So, it is important to define the properties and conditions under which this coupling among different degrees of freedom takes place.
In a recent paper, Blanco-Canosa et al. \cite{prl_fran}. proposed that in single-valence systems, the transition between localized and itinerant electron behavior takes place through a transitional phase in which partial electronic delocalization in the form of cation-clusters in an ionic matrix occurs. In this situation, strong lattice instabilities can be anticipated, due to a purely electronic mechanism \cite{alv2o4_v_molecules}.

In this paper we explore, using ab initio calculations, how the effect of a small (realistic) U/W ratio naturally couples charge, spin and lattice degrees of freedom in a magnetic insulator close to the itinerant-electron limit. Particularly, we will discuss the case of ZnV$_2$O$_4$, whose magnetic and orbital properties are still a matter of debate \cite{motome,chernisov,valenti,mandrus}. We will show that, for small values of U, the most stable structure always consists of V-V dimers along the [011] and [101] directions, that helps to understand the intriguing magnetic structure of the material. This is a dramatic example of a strong electron-lattice coupling due to the partial electronic delocalization in the proximity of the itinerant electron limit.

The electronic structure of the V$^{3+}$ ion (d$^{2}$) in a tetragonally distorted octahedral environment has an orbitally degenerate configuration. It is not yet clear what orbital ordering, if any, could account for the experimentally found magnetic structure of the material \cite{magn_znvo}. Several pictures have appeared recently \cite{chernisov,odps_khomskii,valenti,motome,jackeli}, based on different considerations, that predict various possible orbital orderings. An ``antiferro-orbital" picture has been proposed \cite{motome}, with the full occupation of a d$_{xy}$ orbital at each site and with an alternation of d$_{xz}$ and d$_{yz}$ orbitals along the c-axis. Also, a ``ferro-orbital" ordering of complex orbitals (d$_{xz}$ $\pm$ $i$ d$_{yz}$) \cite{chernisov} and an ``orbital-Peierls" ordering\cite{odps_khomskii} with the orbitals in a d$_{yz}$-d$_{yz}$-d$_{xz}$-d$_{xz}$ pattern along the tetragonally compressed c-axis were suggested. Considering a fully ab initio, all-electron picture, including spin-orbit effects, the most stable solution is an alternating orbital ordering along the [011] and [101] directions of the orbitals with an unquenched orbital angular momentum d$_{xz}$ $\pm$ $i$ d$_{yz}$ \cite{valenti}, such that the orbital angular momentum in every site is antiparallel to the spin moment. However, most of these works rely on the assumption that U/W is large, a purely localized picture, which seems to contradict recent experimental findings  \cite{prl_fran}. In many of these works, the lattice symmetry found in the existing experiments (space group I4$_1$/amd) \cite{tetragonal} was imposed as a rigid constraint. However, in general, electronic structure may have lower symmetry. If this were the case, it would be difficult to detect if corresponding distortions are weak, as in the case of ZnV$_2$O$_4$ (t$_{2g}$ electrons involved). To take into account this possibility, we carried out ab initio electronic structure calculations for ZnV$_2$O$_4$, valid for intermediate values of the Coulomb (Hubbard) repulsion U, and allowing for the lower symmetry, to find the optimal electronic and lattice structure.

We present here full-potential, all-electron, electronic structure
calculations based on the density functional theory (DFT), utilizing the APW+lo
method \cite{sjo}, performed using the WIEN2k software \cite{wien}. For
the structure optimization, we used the GGA (generalized gradient
approximation) in the Perdew-Burke-Ernzerhoff (PBE) scheme \cite{gga} as an exchange-correlation functional.
The geometry optimization was carried out minimizing the forces in the atoms and the total energy of the system, relaxing it from an initial trial state: unequal (dimerized) V-V distances along the [011] and [101] directions.
For the electronic structure calculations we included strong
correlations effects by means of the LDA+U scheme \cite{sic}, where correlation effects are controlled by an effective $U$ ($U_{eff}$= $U-J$), $U$ being the on-site Coulomb repulsion and $J$ the Hund's rule exchange constant. Spin-orbit effects have been
introduced as a second variation using the scalar relativistic
approximation \cite{singh}. All our calculations were fully converged with respect to the parameters used.

ZnV$_2$O$_4$ crystallizes in a distorted cubic spinel structure, where the V atoms form a pyrochlore lattice of corner-sharing tetrahedra. Because of the geometry of the pyrochlore lattice, the AF interactions between the V atoms are highly frustrated, leading to small T$_N$/$\Theta_{CW}$ ratios.

\begin{figure}
\includegraphics[angle=-90,width=\columnwidth]{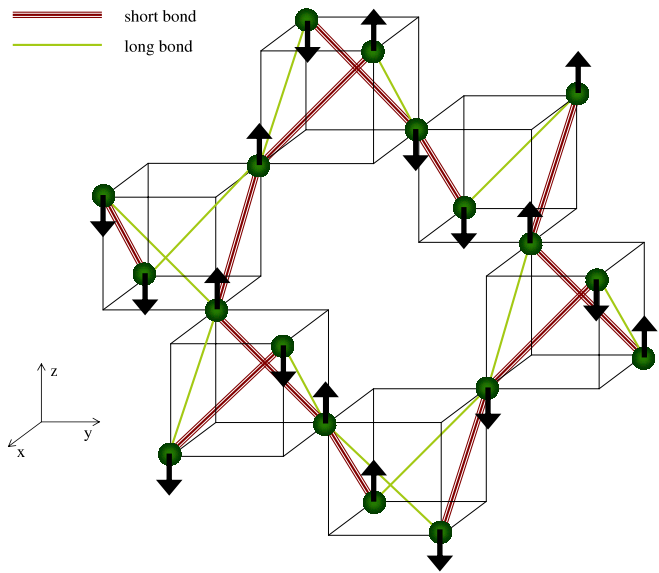}
\caption{(Color online)Schematic representation of the ``dimerized" structure resulting from our ab initio, all-electron calculations for realistic values of U in ZnV$_2$O$_4$. Bold (thin) line represents short (long) in-chain bonds. The magnetic structure is indicated by arrows.
}\label{strfig}
\end{figure}

The material undergoes a magnetic transition and a structural transition to a low-temperature tetragonal phase (c/a$<$ 1) below 50 K \cite{tetragonal}.
The lattice formed by the V atoms can be described as built up by three V-V chains running along the [110], [011] and [101] directions (we use below the cubic setting). The magnetic structure, found by neutron diffraction \cite{magn_znvo}, is AF along the [110] direction (within the ab plane), but along the [101] and [011] (off-plane) directions the spin ordering alternates, the V moments order $\uparrow\uparrow\downarrow\downarrow\uparrow\uparrow$ (see Fig. \ref{strfig}).

In the rest of the paper, for the sake of clarity, the structure derived from the tetragonal distortion described above will be called the ``standard" structure, and the relaxed structure (see below) will be called the ``dimerized" structure, to signal the formation of V-V dimers along the chains.

\begin{figure}
\includegraphics[width=\columnwidth]{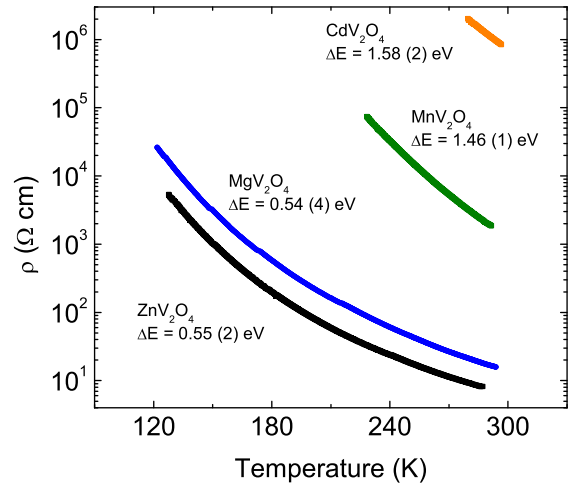}
\caption{(Color online)Temperature dependence of the resistivity in the A$^{2+}$V$_2$O$_4$ series. Small differences in the resistivity for MgV$_2$O$_4$ and ZnV$_2$O$_4$ could be due to the intergrain boundary scattering in the ceramic pellets (their room temperature resistivity differs only by $\simeq$ 8 $\Omega$cm). In any case, their activation energy ($\Delta$E) is practically identical, as it should be, given the similar V-V distance in both compounds.}\label{strfig2}
\end{figure}

As we said before, the anomalous variation found in the pressure dependence of T$_N$ along the series A$^{2+}$V$_2$O$_4$ \cite{prl_fran}, has been interpreted as the consequence of a variation of U/W, which first becomes progressively reduced by pressure, and then collapses close to the itinerant limit. According to this, even though MgV$_2$O$_4$ and ZnV$_2$O$_4$ are still semiconducting, a partial electronic delocalization along the V-V bonds leading to the formation of short-range cation clusters and to a lattice instability can be anticipated. In Fig. \ref{strfig2} we present the experimental resistivity curves for different vanadates in which the divalent cation at the tetrahedral site has been used to tune the V-V distance. The results show how the activation energy decreases as the V-V distance does, and hence the metal-insulator transition is approached from the insulating side. On approaching this point, U/W is reduced progressively \cite{prl_fran} and hence partial delocalization can be anticipated along the V-V bonds. In order 
to check the possibility of a lattice instability due to this electronic effect, we performed a structural optimization, using the experimentally observed magnetic structure ($\uparrow\downarrow\uparrow\downarrow\uparrow\downarrow$ spins in [110] chains, $\uparrow\uparrow\downarrow\downarrow\uparrow\uparrow$ in [101] and [011] directions). As a starting point, we selected an artificially deformed structure that would give rise to V-V dimers (distances along the [101] and [011] chains get short-long-short-long) and let the system relax until forces on the atoms are smaller than 4 mRy/a.u. (small enough to consider the system is relaxed). The system relaxes to a structure away from the ``standard" one, forming chains with an alternation of short-long V-V distances (Fig. \ref{strfig}). The same relaxed structure is obtained starting from different articially deformed structures, including the one with the chains having short-intermedite-long-intermediate V-V distances along the [101] and [011] chains (the structure proposed in Ref. \onlinecite{odps_khomskii}). This shows we have found a stable (its energy is lower than the ``standard" one by 30 meV/V for U$_{eff}$= 0) and reproducible structure. 

The reason why one needs to displace the V atoms from their ``standard" positions is because the latter is a local minimum for the system, as our calculations show (very small forces on the atoms). However, if we move the system slightly out of that structural local minimum, a different local minimum can be found, the one with dimerization shown in Fig. \ref{strfig} with a smaller total energy and hence more stable.

For carrying out this computational experiment, one needs to reduce the symmetry of the compound from the ``standard"  I4$_1$/amd to the P4$_1$2$_1$2 space group, that allows having different V-V distances along the chains and, at the same time, keeping the same magnetic unit cell. The lattice parameters were not optimized, they were taken from Ref. \onlinecite{magn_znvo}. The accuracy of a GGA-based optimization of the lattice parameters is approximately 1\%, which is worse than the experimental resolution. However, the internal positions are much more difficult to obtain experimentally, specially in the case of small distortions as the one we describe here. For that reason, the internal optimization we have carried out computationally becomes necessary.

We have performed a comparison of the total energy of the ``dimerized" structure and the ``standard" one for various values of $U_{eff}$. For this we needed to find the energy minimum for the ``standard" structure. 
The V$^{3+}$:d$^2$ ions in the tetragonal environment have an orbitally degenerate configuration, with the d$_{xy}$ orbital fully occupied and the additional electron in a d$_{xz}$-d$_{yz}$ doublet.
We tried different possible orbital orderings, and the energy minimum, for the ``standard" structure, once spin-orbit coupling is included, is found for the second electron in an l$_z$= $\pm$1 state, with the direction of the orbital angular momentum antiparallel to the spin on each site, which agrees with previous results \cite{valenti}.

\begin{figure}
\includegraphics[width=\columnwidth]{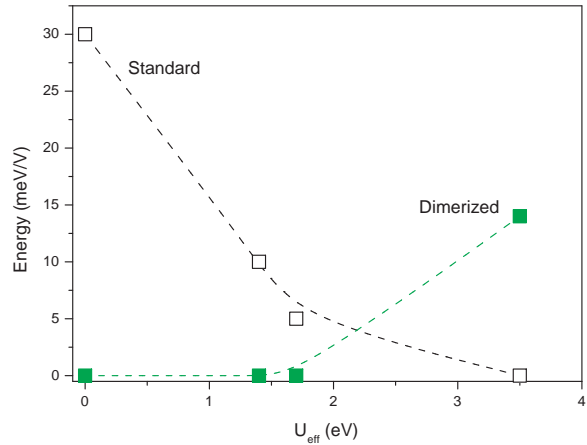}
\caption{(Color online)Total energies (in meV/V) of the two structures as a function of U$_{eff}$. For small values of $U_{eff}$ the ``dimerized" structure proposed in this work is more stable. Note that, for any U$_{eff}$, we take as zero the energy of the most stable structure.}\label{enetab}
\end{figure}

We have compared the total energy of the two structures, the best ``standard" one and the one with dimerization, for different values of U$_{eff}$ (see Fig. \ref{enetab}). For small values of U$_{eff}$ (below about 2.2 eV) the ``dimerized" structure is more stable than the ``standard" one. Such small values of U$_{eff}$ would be consistent with our experimental findings of the system being close to the itinerant  electron limit. It is remarkable that in the ``dimerized" structure, even a GGA calculation is enough to open a tiny gap in the density of states ($\sim$ 0.05 eV). This occurs because in the ``dimerized" structure, spin parallel V atoms move closer to allow two halves of d$_{xz}$ (d$_{yz}$) electrons to fill up a bonding molecular orbital along the [101] ([011]) direction (see Fig. \ref{strfig}). Then the bonding-antibonding splitting opens the gap and stabilizes this ``dimerized" structure, which does not need occurrence of an orbital ordering (see below).
 This is contrary to the case of isostructural spinel MgTi$_2$O$_4$, where the structure along the off-plane chains is short/intermediate/long/intermediate, where no band gap opens within the GGA scheme \cite{mgtio}.

For $U_{eff}$= 1.7 eV, the ``dimerized" structure is more stable than the ``standard" one by 5 meV/V. For a more delocalized case (smaller values of $U_{eff}$ compared to the bandwidth W= 2 eV), the ``dimerized" structure is always the lowest-energy solution. However, if we increase the electron-electron on-site interaction U, i.e. driving the material towards the strongly localized limit, the most stable situation is the ``standard" structure. So, from our calculations we conclude that as U/W is reduced on approaching the itinerant behavior, a strong coupling between charge and lattice degrees of freedom takes place, determining the low-temperature properties of the system. The ``dimerized" structure is consistent with the experimental evidences of lattice instabilities, and the fact that it is stable for small values of U$_{eff}$ confirms our hypotheses about its origin. It is remarkable that no special constraints have been introduced in the calculations, so the specific results obtained for ZnV$_2$O$_4$ may be a manifestation of a general trend which may exist also in other systems close to a localized-itinerant crossover \cite{mazin}.

\begin{table}[h]
\caption{V-V distances (in \AA) along the different directions in the ``standard" (experimental values from Ref. \onlinecite{magn_znvo}) and the ``dimerized" (our calculations) structure of ZnV$_2$O$_4$ and also for the other members of the series AV$_2$O$_4$ in the cubic phase (our experimental results).}\label{distab}
\begin{tabular}{|c|c|c|c|}
\hline
& in-plane & off-plane short & off-plane long\\
\hline
\hline
``standard" & 2.98 & 2.97 & 2.97 \\
``dimerized" & 2.98 & 2.92 & 3.01 \\
CdV$_2$O$_4$ & 3.07 & 3.07 & 3.07 \\
MnV$_2$O$_4$ & 3.01 & 3.01 & 3.01 \\
MgV$_2$O$_4$ & 2.97 & 2.97 & 2.97 \\
\hline
\end{tabular}
\end{table}

In Table \ref{distab} we can see the V-V distances in the ``dimerized" structure and we can compare them with the other members of the series. The shortest V-V distance is below the critical distance for electron itineracy ($\sim$ 2.94 \AA), as estimated by Goodenough for a direct V$^{3+}$-V$^{3+}$ bond across a shared octahedral edge in an oxide \cite{goodenough_oxides}.
This is in agreement with the experimental prediction of the formation of V-V molecular orbitals close to a metal-insulator transition in ZnV$_2$O$_4$ \cite{prl_fran}, and is similar to the case of MgTi$_2$O$_4$, where a tetramerization of the Ti chains has been observed\cite{mgtio}.

As mentioned above, for the ``standard" structure it is possible to stabilize different orbital orderings. Among them, the most stable one has an  unquenched orbital angular momentum of about 0.7 $\mu_B$ per V site antiparallel to the spin moment at each site. However, if we analyze the electronic structure of the ``dimerized" structure, such an orbital ordering is not found. In fact, we do not observe any orbital ordering. The occupations of the levels d$_{xz}$ and d$_{yz}$ are almost identical but the orbital angular momenta are fairly small (about 0.1 $\mu_B$). If we use a basis set with the real combination of orbitals: d$_{xz}$ $\pm$ d$_{yz}$, they are also equally populated. Hence, there is no trace of orbital ordering left once dimerization of the V chains occurs.
The results we present here do not depend on a particular orbital ordering. The dimerization of the structure is caused by a spin-lattice coupling without orbital ordering being involved, the only necessary ingredient is the collapse of U/W in the vicinity of a metal-insulator transition.

In all our calculations, we have assumed that the magnetic structure is the one obtained experimentally \cite{magn_znvo}. But we have also carried out the calculations for different magnetic orderings to try to discern the values of the different exchange constants, assuming there exist an in-plane coupling (J$_{in}$) and an out-of-plane coupling (J$_{out}$). We have used the total energies obtained for various magnetic couplings, with U$_{eff}$= 1.7 eV, fitting them to a Heisenberg model (H= -$\sum_{i,j} J_{ij}S_iS_j$) and estimating the magnetic coupling.
Different results are obtained in the ``dimerized" and ``standard" structures (the latter with an orbital ordering with l$_z$= $\pm$1-orbitals). In the ``dimerized" structure J$_{out}$ can be subdivided in the coupling for the short J$_s$ and for the long bonds, J$_l$. The in-plane coupling remains constant and is strongly AF: J$_{in}$= -16 meV for both structures, mainly due to the singly occupied d$_{xy}$ orbital. Changes occur in the out-of-plane coupling: J$_{out}$= -12 meV for the ``standard" structure (highly frustrated effective AF coupling). For the ``dimerized" structure, J$_s$= 10 meV and J$_l$= -3 meV. The exchange in short bonds is FM, and AF in the long ones. This dimerized structure removes the magnetic frustration and explains the experimentally observed off-plane $\uparrow\uparrow\downarrow\downarrow\uparrow\uparrow$ magnetic structure (see Fig. \ref{strfig}). Our work resolves the difficulty in obtaining the correct magnetic structure met in other approaches \cite{motome,chernisov,jackeli}.

The structure we have obtained is only stable in a limit close to itineracy, when $U$ is comparable to $W$, and hence our conclusions can be applied to ZnV$_2$O$_4$ and not to more localized members of the series, like MnV$_2$O$_4$ and CdV$_2$O$_4$, because they are far from the metal-insulator transition. It is the closeness to the transition what leads to the formation of V-V dimers giving rise to the appearance of molecular orbitals with the subsequent charge delocalization in a dimer in a material that is still semiconducting.

Summarizing, ab initio calculations show that homopolar V-V bonding occurs in some members of the A$^{2+}$V$_2$O$_4$ series. The appearance of this effect is determined by a considerable reduction of U/W, as it has been proposed to occur close to the itinerant-electron limit. Our results prove that charge and lattice degrees of freedom couple strongly in magnetic insulators that approach the itinerant-electron limit. A possible physical picture explaining why there occurs dimerization in the $\uparrow\uparrow\downarrow\downarrow\uparrow\uparrow$ chains close to the itinerant regime is that in this case the exchange interaction resembles double exchange. Shortening of $\uparrow\uparrow$ bonds leads to a larger hopping and to a gain in double exchange energy, stabilizing this spin ordering. In contrast, reduced hopping in longer V-V bonds weakens this tendency, allowing for $\uparrow\downarrow$ ordering in such bonds. The unusual properties of many localized-electron systems that are close to the itinerant crossover should be revisited on the light of the results presented in this work.

The authors thank the CESGA (Centro de Supercomputacion de
Galicia) for the computing facilities and the
Ministerio de Educaci\'on y Ciencia (MEC)  for the financial support through the projects
MAT2006/10027 and HA2006-0119 and also the Xunta de Galicia through the project PXIB20919PR. F.R. also acknowledges MEC for support under program Ram\'on y Cajal. The work of D.Kh. and H.W. was supported by DFG via the project SFB 608, and by the European Project COMEPHS.

\end{document}